\documentstyle[mprocl]{article}

\def\be{\begin{equation}}
\def\ee{\end{equation}}
\def\bea{\begin{eqnarray}}
\def\eea{\end{eqnarray}}

\begin{document}

\title{HOLOGRAPHIC BOUND FROM SECOND LAW}

\author{JACOB D. BEKENSTEIN}

\address{\it The Racah Institute of Physics, Hebrew University of
Jerusalem,\\ Givat Ram, Jerusalem 91904, Israel\\E-mail:
bekenste@vms.huji.ac.il}   

\maketitle\abstracts{ 
The holographic bound that the entropy (log of number of quantum states) of a
system is bounded from above by a quarter of the area of a circumscribing
surface measured in Planck areas is widely regarded a desideratum of any
fundamental theory, but some exceptions occur. By suitable black hole gedanken
experiments I show that the bound follows from the generalized second law for
two broad classes of isolated systems: generic weakly gravitating systems
composed of many elementary particles, and quiescent, nonrotating strongly
gravitating configurations well above Planck mass.  These justify an early
claim by Susskind.}

\section{What is the holographic bound ?}
\noindent
The holographic principle, first enunciated by 't Hooft,~\cite{thooft} claims
the physical equivalence between pairs of physical theories.  One member of a
pair, T1, describes a bulk system ${\cal U}$ in a spacetime; the allied
theory, T2, describes a boundary of that spacetime.  For instance, string
theory in the $D=10$ spacetime $AdS_5\times S^5$ is known to be equivalent to
a supersymmetric gauge theory on the boundary.  Some more examples are known,
mostly in $D>4$.  Many regard the holographic principle as a criterion for
good physics.   

An obvious consistency requirement on the holographic principle is
that the boundary of any system should be able to encode as much information as
required to catalogue the quantum states of the bulk system:  T2
should allow at least as many quantum states to reside on the boundary as T1
allows for the bulk (otherwise, so to speak, T2 does not know enough to
be equivalent to T1).  Suppose we go to $D=4$ and take the
logarithm of this inequality.  On the one hand we get the entropy
${\cal S}$ of ${\cal U}$ and its gravitational field; on the other we get the
entropy of the boundary, which analogy with black hole entropy suggests to
quantify by one quarter of its 2-D area ${\cal A}$ in Planck units.  Thus we
have the guess (the holographic bound~\cite{thooft,susskind} here denoted HB;
Planck units used throughout !)
\begin{equation}
{\cal S}\leq {\cal A}/4
\label{bound}
\end{equation}
How do we know that this bound is true ?

As support for the HB, Susskind~\cite{susskind} described a {\it gedanken\/}
experiment in which a system violating the HB is forced to collapse to a black
hole by adding to it extra entropy--free matter. Susskind interprets the
ensuing apparent violation of the generalized second law (GSL) as evidence that
the envisaged system cannot really exist.  In the clearer reformulation of
Wald,~\cite{wald} one imagines the system as a spherically symmetric one of
radius $R$, energy $E$ (with $R>2E$, of course) and  entropy ${\cal S}$ which
violates the HB: ${\cal S}>\pi R^2$.  A spherically symmetric and concentric
shell of mass $R/2-E$ is dropped on the system; by Birkhoff's theorem the total
mass is now $R/2$.  {\it If\/} the outermost surface of the shell reaches
Schwarzschild radial coordinate $r=R$, the system becomes a black hole of
radius $R$ and entropy $S_{BH}=\pi  R^2$, which is lower than the original
entropy ${\cal S}$ !  Susskind would conclude from this that, contrary to the
assumption, the HB cannot be violated for ${\cal U}$.  However, Wald points
out an alternative conclusion: if ${\cal S}>\pi R^2$ the outcome of the
process is not a black hole, e.g. the shell could bounce, or a naked
singularity appear.  Thus the Susskind argument is inconclusive.

Below I review two variations~\cite{bek00} on Susskind's argument which provide
proof of the  HB for a broad class of weakly gravitating systems and a broad
class of strongly gravitating systems.  I should mention that there do exist
systems which violate the HB, e.g. a large enough spherical section of a flat
Friedmann universe, or a collapsing sphere inside its gravitational
radius.~\cite{bousso}

\section{Systems with weak self--gravity}

Consider first a weakly self--gravitating system ${\cal U}$ which may have
arbitrary structure and constitution, but require it to be isolated to avoid
the problem of the large section of the universe.  Let ${\cal U}$'s proper
mass--energy be $E$ and its entropy ${\cal S}$.  Further, imagine ${\cal U}$ is
enclosed in a snug spherical box of radius $R$ concentric with ${\cal U}$'s
center of mass (c.m.) and let the box be made of entropy--less  material. 
Any essential mass of the box is to be included in $E$.   Weakly
self--gravitating means $R\gg E$.  Assume in addition that $R\gg 1/E$:  the
system is far from being an elementary particle, and thus large compared to its
own Compton length. These assumptions together imply that $R\gg 1$ (we always
deal with systems large on Planck length scale).  Now imagine dropping the
sphere freely into a Schwarzschild black hole of mass $M=R^2(8E)^{-1}$
(assuming, of course that it is possible to find black holes of any mass).  Now
because  $R\gg E$,  $M\gg E$ and $2M\gg R$.    Thus ${\cal U}$ constitutes a
small perturbation on the black hole and is very small compared to it; there is 
thus no reason why the black hole should be destroyed by its infall or why
${\cal U}$ should be torn up by tidal forces.  

Now if ${\cal U}$ falls from rest at some point far from the hole, its  energy
as measured at infinity is $< E$; writing ${\cal E}$ for the energy lost by
${\cal U}$ and black hole to radiation (see below) during the infall, we see
that the hole increases its mass by $<E-{\cal E}$. The black hole's initial
entropy, $S_{BH} = 4\pi M^2$, thus increases by $\Delta S_{BH}<8\pi M(E-{\cal
E})[1+{\scriptstyle 1\over\scriptstyle 2}(E-{\cal E})/M]$.  By the GSL, $\Delta
S_{BH}$ plus the radiations entropy, $S_{\rm rad}$, must be at least as large
as ${\cal S}$, the entropy which disappears from sight.  We thus have
\begin{equation}
{\cal S}< 8\pi M(E-{\cal E})[1+{\scriptstyle 1\over\scriptstyle
2}(E-{\cal E})/M] + S_{\rm rad}.
\label{raw_bound}
\end{equation}

Now the energy lost by the infalling ${\cal U}$ to gravitational radiation  is
known~\cite{davis} to be of $O(E^2/M)$, here reducing to
$O\big(E(E/R)^2\big)$.  Because $R\gg E$ we may neglect this
contribution to ${\cal E}$ in the last inequality in comparison with the much
bigger $E$.   The other contribution to ${\cal E}$ is by Hawking radiance of
the hole.  One can approximate the power so emitted by that of a sphere with
radius $2M$  radiating according to the Stefan--Boltzmann law at the Hawking
temperature $T_H= (8\pi M)^{-1}$.  Thus the rate of black hole mass change is  
\begin{equation}
\dot M \approx - {N\over 15360 \pi M^2}, 
\label{Mdot}
\end{equation}
where $N$ denotes the effective number of radiated species. If  ${\,\cal U}$
was dropped at Schwarzschild time $t=0$ when its center of mass was at
$r=\alpha 2M\ (\alpha\gg 1)$, then integration of the geodesic equation
describing ${\cal U}$'s c.m. motion in the Schwarzschild metric shows that
the  {\it bottom\/} of the sphere reaches the horizon $r=2M$ (and the sphere's
center reaches {\it proper\/} height $R$ above it), and the infall is finished
for all practical purposes, at  time 
\begin{equation}
t \approx 4M\left[{\scriptstyle 1\over \scriptstyle
3}\alpha^{3/2} -{\scriptstyle 1\over \scriptstyle
4}\ln(R/4M)+O(\alpha^{1/2})\right].
\label{time}
\end{equation} 
The textbook statement that a particle
takes an infinite $t$ time to reach the horizon is recovered from the
$R\rightarrow 0$ limit here; but obviously that eternity is only
logarithmically big ! 

The Hawking energy emitted during the infall to the black hole of mass
$M=R^2(8E)^{-1}$ is thus ${\cal E}=|\dot M|\times t \approx N(480\pi
R^2)^{-1}\left[{\scriptstyle 1\over \scriptstyle 3}\alpha^{3/2}-{\scriptstyle
1\over \scriptstyle 4}\ln(2E/R) \right] E.$  Because $R\gg 1$, the factor
$N(480\pi R^2)^{-1}$ is very small compared to unity given that in nature
$N=O(10)$.  In fact this small factor easily outbalances any moderately large
$\alpha^{3/2}$ and the logarithmic factor which is never big.  Thus ${\cal E}
\ll E$, and hence it is justified to drop ${\cal E}$ everywhere in
(\ref{raw_bound}).  Also, because $E\ll R\ll 2M$ we can drop the
$E/M$ correction in Eq.~(\ref{raw_bound}).  The last two
inequalities also justify {\it a posteriori\/} our use of a Schwarzschild
metric of mass $M$ for calculating the time interval.

What about $S_{\rm rad}$ in Eq.~(\ref{raw_bound}) ?  The infall gravitational
radiation is coherent and so should carry negligible entropy, particularly
since its energy is small.  The Hawking radiance entropy is $\sim {\cal
E}/T_H=8\pi M{\cal E}$, obviously negligible compared to the dominant term,
$8\pi ME$, in Eq.~(\ref{raw_bound}).  Putting $M=R^2(8{\cal E})^{-1}$ in that
formula we get ${\cal S}< \pi R^2$, which is precisely the HB,
Eq.~(\ref{bound}). 

The above treatment presumes that radiation pressure from the Hawking's
radiance does not  prevent  ${\cal U}$ from reaching the horizon and does not
drastically prolong the infall time estimated in Eq.~(\ref{time}).   Are these
suppositions true ?  The momentum {\it flux\/} from Hawking radiance at radial
coordinate
$r$ is obviously $|\dot M|(4\pi r^2)^{-1}$.  Assuming that all the radiation
hitting the sphere is absorbed, the rate at which it picks up momentum from the
radiation, in its own rest frame (four velocity $u^\beta$), is at least this
flux times the sphere's crossection [which is $O(\pi R^2)$], times
$(dt/d\tau)^2$, where $\tau$ is the sphere's proper time (if the sphere
reflects radiation, one must multiply this result by a factor between 1 and 2
to account for backscattering).  One factor $dt/d\tau$ accounts for the
blueshift of the momenta of the Hawking quanta perceived in the falling
sphere's frame;  the second corrects for the faster arrival of quanta due to
the time dilation and gravitational redshift. Since the sphere falls from
$r\gg 2M$, we have
$dt/d\tau\approx(1-2M/r)^{-1}$.  Putting all the factors together with
Eq.~(\ref{Mdot}) and dividing by $E$,  we find the acceleration of the falling
sphere measured in its own frame (its acceleration scalar):
\begin{equation}
a\approx{N  R^2\over 61440  M^2 Er^2}{1\over (1-2M/r)^2}
\label{acceleration}
\end{equation}

Is this big or small ? The quantity to compare it with is the acceleration 
scalar of a stationary point at radial coordinate $r$, $g
=Mr^{-2}(1-2M/r)^{-1/2}$:
\begin{equation}
{a\over g}={N/E\over 7680  R}\left[{R\over
2M(1-2M/r)^{1/2}}\right]^3.
\label{acceleration_ratio}
\end{equation}
Since we assumed $R\gg 1/E$ and have $R\ll 2M$ in our {\it gedanken\/}
experiment, it is apparent that if the sphere is not very close to the black
hole's horizon, the radiation pressure acceleration is negligible negligible on
the natural scale $g$.  And because $R\ll 2M$, the factor in square brackets
in Eq.~(\ref{acceleration_ratio}) only grows to 2 when the bottom of the
sphere touches the horizon.  Hence the radiation pressure deceleration is
totally negligible throughout.  We conclude that ${\cal U}$'s c.m. does
move accurately on a timelike geodesic all the way down to the horizon, as has
been assumed all along.  Smallness of $a/g$ also means that it is unnecessary
to correct for quantum buoyancy effects, as is the case when the system is
suspended.\cite{bek99}

\section{Systems with strong self--gravity}
 
\noindent
The argument in Sec.~2 fails when the system's self--gravity becomes
strong, say, $R< 20 E$ because in this case, $M=R^2(8E)^{-1}$ implies that
$2M<5 R$ so we cannot be sure that ${\cal U}$ is not tidally torn up by the
black hole.  In addition we see that $M< 50E$ so that ${\cal U}$ is a
significant perturbation on the hole.  This means we cannot assume it will
fall on a geodesic of the background metric, or even that it will not cause a
singularity on the horizon.  We thus change our strategy; we now employ a
small black hole, denoted ${\cal H}$ below, whose task is to  catalyze the
conversion of the strongly gravitating system, now denoted ${\cal V}$, into a
single large black hole.

As before we assume ${\cal V}$ is isolated, which means its exterior geometry
is asymptotically flat; this makes it straightforward to define its total
energy $E$.  To simplify the deductions we assume that ${\cal V}$ is quiescent
and nonrotating, i.e. that its geometry is nearly static.   And finally we
shall assume that $E\gg {\rm Max}(\mu^{-1}, 10^3 \surd N)$ where $\mu$ is the
mass of the lightest massive elementary constituent of ${\cal V}$ and
$N$, again, is the number of species in radiation.   These restrictions will
guarantee compatibility of the bounds that we shall assume presently on 
${\cal H}$'s mass.  Note that ${\cal V}$ is required to be very massive on
Planck scale.

Now, it is pretty clear that the area ${\cal A}$ of a 2-D surface
enclosing ${\cal V}$ must exceed $4\pi(2E)^2$ for otherwise ${\cal V}$ would
already be a black hole (there are pathological surfaces which can be smaller;
see Ref.~4 for a cleaner definition).  And it must exceed it substantially;
otherwise ${\cal V}$ would be unstable against black hole formation, i.e., not
quiescent.  We thus take it that ${\cal A}$ is substantially greater than
$16\pi E^2$.

Let us enclose ${\cal V}$ in a roomy quasispherical box concentric with its
c.m. whose walls are robust enough to trap all radiation that will be
produced, save for gravitational waves which penetrate almost every barrier. 
We take the inner radius of the box to be $r\approx 2\cdot 10^2 E$ (in the
sense of approximate Schwarzschild coordinates).   ${\cal V}$ being strongly
gravitating, it evidently occupies only a small central region of the box;
therefore, most of the box volume is close to flat spacetime.  The box is
likely to be massive, and so to cause a shift in the standards of time and
energy within it as compared with the outside world.  All statements about
these quantities refer to the interior.  

Let ${\cal H}$'s mass $m$ be restricted by ${\rm Max}(\mu^{-1}, 10 N^{1/5}
E^{3/5}) \ll m\ll E$.  The upper and lower bound here are consistent by virtue
of the constraints on $E$ we adopted.  We drop the small Schwarzschild black
hole ${\cal H}$ from rest from within the box and near its wall  and with no
angular momentum.      Because ${\cal V}$ is strongly gravitating but
non--rotating, it will be almost spherical and so we may roughly approximate
its exterior metric by Schwarzschild's.  Then by Eq.~(\ref{time}),
${\cal H}$ takes an (infall) time $t_f \sim 10^3 E$ to reach ${\cal V}$.  This
does not mean ${\cal H}$ stops within ${\cal V}$; the latter may be
tenuous enough to allow ${\cal H}$ to cross it, rise within the box to near the
wall, and turn around for another such cycle.  In one extreme case the black
hole is trapped in ${\cal V}$ after one or two such cycles; in the other the
trapping time is maximal, $t_t$, and corresponds to the number of passes
through ${\cal V}$ that ${\cal H}$ must make in order for its crossection
$16\pi m^2$ to sweep through the whole volume of ${\cal V}$, thus insuring a
hard collision with the more massive system.  Because the linear extent of
${\cal V}$ is of order
$({\cal A}/4\pi)^{1/2}$, it takes some $(E/m)^2$ passes to do this, assuming,
in harmony with the fact we did not take ${\cal V}$ to be spherically
symmetric, that each pass through it is in a different direction.  Thus
$t_t\sim 10^3(E/m)^2 E$.

Once trapped in ${\cal V}$, the little hole begins to digest its host.   Since
${\cal V}$ is strongly self--gravitating, ${\cal H}$ will at first move
at relativistic speed so that accretion may be inefficient.  But collisions
with ${\cal V}$'s components will slow it down (which is entirely possible if
it settles down deep in the gravitational well of ${\cal V}$), and allow the
accretion to speed up.  Aided by its tidal field, ${\cal H}$ can tear up parts
of ${\cal V}$ to smaller pieces, down to the scale of elementary
constituents.  Since the largest Compton length of these is smaller than
${\cal H}$'s size ($m\gg \mu^{-1}$), the hole can swallow anything that comes
within its reach.  The duration of this digestion stage obviously depends on
details about ${\cal V}$ but there should be an upper bound to it,  $t_d$,
which should be a  function of only the important scales of the problem $E$
and $m$.  Were $m$ comparable with $E$, the digestion stage would obviously be
over in the crossing time $\sim E$. For small $m/E$ the accretion
effectiveness should scale as ${\cal H}$'s crossection $m^2$.  Thus on
dimensional grounds we guess $t_d\sim (E/m)^2 E$.   Towards the end of the
digestion the joint system ${\cal V}+{\cal H}$ is likely to evolve rapidly. 
Fragments that get ejected from ${\cal V}$  as it is being swallowed remain
trapped in the box,  and will eventually fall back onto ${\cal H}$, be broken
up and swallowed.  This last stage should last a time
$\sim E$ because it obviously involves an instability, instabilities grow
exponentially, and our two scales, $E$ and $m$, are merging as the black hole
ingests all of ${\cal V}$.  We thus see that the digestion is likely  briefer
than the trapping, no matter how big $N$.

But will not ${\cal H}$ Hawking evaporate before it consumes ${\cal V}$ ?  The
Hawking evaporation timescale of the initial black hole is, according to
Eq.~(\ref{Mdot}), $t_H\approx 5120\pi m^3/N$.  But since we assume $m\gg 10
N^{1/5} E^{3/5}$, it follows that $t_H \gg 10^9 (E/m)^2 E$, so over the course
of the trapping and digestion times, ${\cal H}$ hardly loses mass by Hawking
emission.  Of course as ${\cal H}$ gets more massive by accretion, the Hawking
emission weakens even further:  Hawking radiance is negligible in our {\it
gedanken\/} experiment.

Obviously the  accretion will generate heat, and radiation (e.g.
electromagnetic but little gravitational because of its poor coupling to
matter) will thus leak out of ${\cal V}$ into the box, which we have assumed
can trap the nongravitational quanta. As it nears its end, ${\cal V}$ should
approach  spherical symmetry because it is nonrotating, and the elastic forces
that could keep it aspherical will succumb to gravitation as the black hole
gnaws its way through it.  Finally, as the black hole finishes its meal, it
recovers its original Schwarzschild form, and establishes thermodynamic
equilibrium with the radiation filling the box.  In the presence of a
spherically symmetric system at its center, the box can be regarded as
perfectly spherical and thus gravitationally irrelevant (apart from the
gravitational redshift it induces inside it).

How long does it take for the cavity to reach  equilibrium with ${\cal H}$ ? 
An upper bound on this equilibration time $t_{eq}$ is set by the time the
final hole would take to fill the cavity just with Hawking radiation (we saw
that the  evaporation time of the faster radiating initial hole is long
compared to the digestion time which is also the heating time). It shall
transpire that the final hole has a mass very near $E$.  Thus its Hawking
temperature (and the temperature of the equilibrated radiation) is $T=(8\pi
E)^{-1}$.  Using the volume $(4\pi/3)(2\cdot 10^2 E)^3$ for the box's interior
and Boltzmann's energy density for black body radiation, we compute the energy
of radiation trapped in the box as ${\cal E}\approx 55NE^{-1}$.   Dividing
this by the Hawking power (\ref{Mdot}) with $M\rightarrow E$ gives $t_{eq}<
3\cdot 10^6 E$.  Since we assume $E\gg m$, $t_{eq}$ may even be shorter than
$t_d$ or
$t_t$.  

It might be claimed that not $t_{eq}$, but the relaxation time $t_r$ for
the radiation in the box is the informative timescale: if the radiation and
hole are slightly out of equilibrium, how long do they take to reach it or
return to it ? Intuitively this time should be of order of the times for
absorption and reemission of a typical quantum of radiation by ${\cal H}$; in
or near equilibrium these two equivalent.  Now a typical quantum not very
close to ${\cal H}$ will be absorbed in its next crossing of the box with
probability roughly equal to the solid angle subtended by the hole at the
quantum's starting point divided by $4\pi$.  This is about $(10^{-2})^2$ for
our cavity.  Dividing the light crossing time $2\cdot 10^2 E$ by this
probability gives the estimate $t_r\sim 2\cdot 10^6 E$ which is very like the
upper bound on $t_{eq}$.  Therefore, ${\cal H}$ comes into precise
thermodynamic equilibrium with the cavity's radiation in a time no longer than
it took to digest ${\cal V}$.

Making use twice of our assumption $E\gg 10^3\surd N$ in our previous
expression for ${\cal E}$, we get that the equilibrium radiation energy in the
box is ${\cal E}\approx 55NE^{-1}\ll 10^{-4} E$.  And since $m\ll E$ the energy
of the final black hole is very near $E$.  But what about the energy ${\cal
E}_g$ carried away by Hawking gravitons ?  In its initial stages the black
hole's power in gravitons is given by Eq.~(\ref{Mdot}) with $N=1$ and
$M\rightarrow m$.  Multiplying this by, say, the trapping time $t_t$ gives
${\cal E}_g \sim 0.02 E^3 m^{-4}\ll 2\cdot 10^{-7} E N^{-1}$ in view of our
restriction $m\gg 10 N^{1/5} E^{3/5}$ and $E\gg m$.  In its latter stages the
hole's graviton power is again (\ref{Mdot}) with $N=1$ but this time with
$M\rightarrow E$.  Over the relaxation time $t_r$ this gives energy ${\cal
E}_g \sim 10^2 E^{-1}\ll 10^{-4} E N^{-1}$, where we have twice used our
assumption $E\gg 10^3 \surd N$.  In summary, the energy lost to gravitons from
the dropping of ${\cal H}$ into ${\cal V}$ through the equilibration of the
final hole with the cavity is ${\cal E}_g\ll10^{-4} E N^{-1}$.  The upshot of
this paragraph is that it was self--consistent to assume that in its final
state ${\cal H}$ has energy very nearly $E$ and Hawking temperature very nearly
$(8\pi E)^{-1}$.

We now draw up the balance of entropy.  Originally we had entropy ${\cal
S}$ in ${\cal V}$ and $S_{BH}=4\pi m^2$ in the initial ${\cal H}$.  Just after
its equilibration and  relaxation, the cavity radiation has entropy
${\scriptstyle 4\over\scriptstyle 3}{\cal E} T^{-1}={\scriptstyle
32\over\scriptstyle 3}\pi E{\cal E}$ (the ${\scriptstyle 4\over\scriptstyle 3}$
is well known from Boltzmann's formulae for black body radiation).  Hawking
gravitons have carried away entropy $S_g\sim {\cal E}_g T_H^{-1}\ll 2\cdot
10^{-7}\cdot 8\pi mE N^{-1}$ in the digestion stage and  $S_g\sim {\cal E}_g
T^{-1}\ll  10^{-4}\cdot 8\pi E^2 N^{-1}$ in the equilibration
stage.  The final ${\cal H}$ has entropy $S'_{BH}=4\pi(E+m-{\cal E}-{\cal
E}_g)^2$.  Applying the GSL and rearranging terms gives us
\begin{equation}
{\cal S}<4\pi E^2\left[1+{2m\over E}+{\scriptstyle 2\over\scriptstyle 3}{{\cal
E}\over E}+{2{\cal E}_g\over E}+{S_g\over 4\pi E^2}+\cdots\ \right]
\label{GSL}
\end{equation}           
where the ellipsis denotes second order terms like $m{\cal E} E^{-2}$ or ${\cal
E}_g{}^2 E^{-2}$.  Let us look at the corrections to ``1'' in the square
brackets.  We have just seen that  in the early stages of ${\cal V}$'s
digestion ${\cal E}_g E^{-1}\ll 2\cdot 10^{-7} N^{-1}$ and $S_g(4\pi
E^2)^{-1}\ll 4\cdot 10^{-7}(m/E) N^{-1}$, while in the latter stages ${\cal
E}_g E^{-1}\ll 10^{-4}N^{-1}$ and $S_g(4\pi E^2)^{-1}\ll 2\cdot 10^{-4}
N^{-1}$; in addition, ${\cal E}E^{-1}\ll 10^{-4}$.  Hence the corrections to
``1'' are tiny (we assumed all along that $m\ll E$).  On the other hand, we
have pointed out that the area ${\cal A}$ of a 2-D surface surrounding the
original system ${\cal V}$ must substantially exceed $16\pi E^2$.  It follows
from inequality~(\ref{GSL}) that ${\cal S}$ of the original system obeyed the
holographic bound (\ref{bound}). 

\section{Summary and Caveats}
\noindent
We have thus vindicated Susskind's idea that the GSL can serve as the basis of
a proof of the HB, albeit a proof for restricted classes of systems. The
present account leaves out a number of details treated in the original
paper.~\cite{bek00}  Beyond that one can imagine situations which would
sidestep the proof in Sec.~3.  For example, were ${\cal V}$ spherically
symmetric save for a straight ``tunnel'' though its center, then by dropping
${\cal H}$ down the tunnel one could arrange for ${\cal H}$ to go back and
forth without ever colliding with anything in ${\cal V}$.  This evidently can
be fixed by dropping ${\cal H}$ in a generic direction.  Or one could fear
that as a result of random photon emission by the half eaten ${\cal V}$, the
final ${\cal H}$ could be left with net momentum with respect to the box, so
that it could collide with and destroy its wall.  Again, a detailed
calculation shows the timescale for the envisaged motion is much longer than
$t_d$ or $t_r$, so if the experiment is done as soon as possible, no problem
occurs.   

\vspace*{-9pt}

\section*{Acknowledgments}
This research was supported by the Hebrew University's Intramural Research
Fund. I thank M. Milgrom and R. Bousso for enlightening comments.

\eject

\end{document}